\documentclass[11pt]{article}
 
\title{Convergence between Categorical Representations of Reeb Space and Mapper\footnote{This work was partially support by NSF IIS-1513616.}}

\author{
Elizabeth Munch\\
Dept.~of Mathematics and Statistics, University at Albany -- SUNY\\
Bei Wang\\
 Scientific Computing and Imaging Institute, University of Utah
}

\date{}

\usepackage{wrapfig}
\usepackage{tikz-cd}
\usepackage{bbm}
\usepackage{amsfonts}
\usepackage{amsthm}
\usepackage{amsmath}

\usepackage[paperwidth=8.5in, paperheight=11in]{geometry}

\numberwithin{equation}{section}
\newtheorem{theorem}[equation]{Theorem}

\newtheorem{lemma}[equation]{Lemma}

\newtheorem{definition}[equation]{Definition}
\newtheorem{corollary}[equation]{Corollary}

\newcommand{\Rcat}{\mathbf{R}}

\newcommand{\Open}{\mathbf{Open}}
\newcommand{\Set}{\mathbf{Set}}
\newcommand{\Vect}{\mathbf{Vect}}

\newcommand{\Top}{\mathbf{Top}}
\newcommand{\Cell}{\mathbf{Cell}}

\newcommand{\colim}{\mathop\mathrm{colim}}
\newcommand{\PCat}{\mathbf{P}}

\newcommand{\Rtop}{\mathbb{R}\textrm{-}\mathbf{Top}}
\newcommand{\Rdtop}{\mathbb{R}^d\textrm{-}\mathbf{Top}}
\newcommand{\Rtopc}{\Rtop^\mathrm{c}}

\newcommand{\Reeb}{\mathbf{Reeb}}

\newcommand{\Cshc}{\mathbf{Csh}^c}

\newcommand{\Nerve}{\mm {\mathrm{Nrv}}}

\newcommand{\inv}{^{-1}}
\newcommand{\e}{\varepsilon}
\renewcommand{\phi}{\varphi}

\newcommand{\diam}{\mathrm{diam}}
\newcommand{\res}{\mathrm{res}}
\newcommand{\image}{\mathrm{image}}
\newcommand{\op}{\mathrm{op}}
\newcommand{\Nrv}{\mathrm{Nrv}}

\newcommand {\mm}[1] {\ifmmode{#1}\else{\mbox{\(#1\)}}\fi}

\newcommand{\Rspace}        {\mm{{\mathbb R}}}

\newcommand{\Xspace}        {\mm{{\mathbb X}}}
\newcommand{\Yspace}        {\mm{{\mathbb Y}}}

\newcommand{\Ucal}{\mathcal{U}}

\newcommand{\R}{\mathbb{R}}

\newcommand{\X}{{\mathbb X}}
\newcommand{\Y}{{\mathbb Y}}

\newcommand{\CC}{\mathcal{C}}
\newcommand{\DD}{\mathcal{D}}
\newcommand{\FF}{\mathcal{F}}
\newcommand{\GG}{\mathcal{G}}

\newcommand{\MM}{\mathcal{M}}
\newcommand{\PP}{\mathcal{P}}
\newcommand{\RR}{\mathcal{R}}
\renewcommand{\SS}{\mathcal{S}}

\newcommand{\UU}{\mathcal{U}}

\begin{document}

\maketitle

\begin{abstract}
The Reeb space, which generalizes the notion of a Reeb graph, is one of the few tools in topological data analysis and visualization suitable for the study of multivariate scientific datasets. 
First introduced by Edelsbrunner et al., it compresses the components of the level sets of a multivariate mapping and obtains a summary representation of their relationships. 
A related construction called mapper, and a special case of the mapper construction called the Joint Contour Net have been shown to be effective in visual analytics. 
Mapper and JCN are intuitively regarded as discrete approximations of the Reeb space, however without formal proofs or approximation guarantees. 
An open question has been proposed by Dey et al.~as to whether the mapper construction converges to the Reeb space in the limit. 

In this paper, we are interested in developing the theoretical understanding of the relationship between the Reeb space and its discrete approximations to support its use in practical data analysis. Using tools from category theory, we formally prove the convergence between the Reeb space and mapper in terms of an interleaving distance between their categorical representations. 
Given a sequence of refined discretizations, we prove that these approximations converge to the Reeb space in the interleaving distance; this also helps to quantify the approximation quality of the discretization at a fixed resolution.

\end{abstract}

\section{Introduction}
\label{sec:introduction}

\subparagraph*{Motivation and prior work.} Multivariate datasets arise in many scientific applications, ranging from oceanography to astrophysics, from chemistry to meteorology, from nuclear engineering to molecular dynamics. 
Consider, for example, combustion or climate simulations where multiple physical measurements (e.g.~temperature and pressure) or concentrations of chemical species are computed simultaneously.
We model these variables mathematically as multiple continuous, real-valued functions defined on a shared domain, which constitute a multivariate mapping 
$f:  \Xspace \to  \Rspace^d$,
also known as a multi-field. 
We are interested in understanding the relationships between these real-valued functions, and more generally, in developing   
efficient and effective tools for their analysis and visualization.

Recently, topological methods have been developed to support the analysis and visualization of scalar field data with widespread applicability. In particular, a great deal of work for scalar topological analysis has been focused on computing the Reeb graph~\cite{Reeb1946}. The Reeb graph contracts each contour (i.e.~component of a level set) of a real-valued function to a single point and uses a graph representation to summarize the connections between these contours. When the domain is simply connected, this construction forms a contour tree, 
which has been shown to be effective in many applications including data simplification and exploratory visualization~\cite{CarrSnoeyinkPanne2010}.   
From a computational perspective, both randomized~\cite{HarveyWangWenger2010} and deterministic~\cite{Parsa2012} algorithms exist that compute the Reeb graph for a function defined on a simplicial complex $K$ in time $O(m\log{m})$, where $m$ is the total number of vertices, edges and triangles in $K$. 
Recent work by de Silva et al.~\cite{SilvaMunchPatel2014} has shown that the data of a Reeb graph can be stored in a category-theoretic object called a cosheaf, which opens the way for defining a metric for Reeb graphs known as the interleaving distance.
The idea of utilizing a cosheaf over a simplicial complex has also been previously investigated, in particular in the work of Curry \cite{Curry2014}.

Unlike for real-valued functions, very few tools exist for studying multivariate data topologically as the situation becomes much more complicated.
The most notable examples of these tools are the Jacobi set~\cite{EdelsbrunnerHarer2002} and the Reeb space~\cite{EdelsbrunnerHarerPatel2008}.  
The Jacobi set analyzes the critical points of a real-valued function restricted to the intersection of the level sets of other functions.
On the other hand, the Reeb space, a generalization of the Reeb graph, compresses the components of the level sets of the multivariate mapping (i.e.~$f^{-1}(c)$, for $c \in \Rspace^d$) and obtains a summary representation of their relationships. 
These two concepts are shown to be related as the image of the Jacobi sets under the mapping corresponds to certain singularities in the Reeb space.  
An algorithm has been described by Edelsbrunner et al.~\cite{EdelsbrunnerHarerPatel2008} 
to construct the Reeb space of a generic piecewise-linear (PL), $\R^d$-valued mapping defined on a combinatorial manifold up to dimension $4$. 
Let $n$ be the number of $(d-1)$-simplices in the combinatorial manifold. 
Assuming $d$ is a constant, the running time of the algorithm is $O(n^d)$, polynomial in $n$~\cite{Patel2010}. 

A related construction called mapper~\cite{SinghMemoliCarlsson2007} takes as input a  multivariate mapping and produces a summary of the data by using a cover of the range space of the mapping. 
Such a summary converts the mapping with a fixed cover into a simplicial complex for efficient computation, manipulation, and exploration~\cite{LumSinghLehman2013, NicolauaLevinebCarlsson2011}. 
When the mapping is a real-valued function (i.e.~$d=1$) and the cover consists of a collection of open intervals, it is stated without proof that the mapper constrcution recovers the Reeb graph precisely as the scale of the cover goes to zero~\cite{SinghMemoliCarlsson2007}. A similar combinatorial idea has also been explored with the $\alpha$-Reeb graph~\cite{ChazalSun2014}, which is another relaxed notion of a Reeb graph produced by a cover of the range space consisting of open intervals of length at most $\alpha$. 
Recently, Dey et al.~\cite{DeyMemoliWang2015} extended mapper to its multiscale version by considering a hierarchical family of covers and the maps between them. At the end of their exposition, the authors raised an open question in understanding the continuous object that the mapper construction converges to as the scale of the cover goes to zero, in particular, whether the mapper construction converges to the Reeb space. 
In addition, Carr and Duke~\cite{CarrDuke2014} introduced a special case of mapper called the Joint Contour Net (JCN) together with its efficient computation, for a PL mapping defined over a simplicial mesh involving an arbitrary number of real-valued functions. 
Based on a cover of the range space using $d$-dimensional intervals, the JCN quantizes the variation of multiple variables simultaneously by considering connected components of interval regions (i.e.~$f^{-1}(a, b)$) instead of the connected components of level sets (i.e.~$f^{-1}(c)$). 
It can be computed in time $O(km \alpha(km))$, where $m$ is the size of the input mesh, $k$ is the total number of quantized interval regions, and $\alpha$ is the slow-growing inverse Ackermann function~\cite{CarrDuke2014}. 
The authors stated that the JCN can be considered as a discrete approximation that converges in the limit to the Reeb space~\cite{CarrDuke2014}, although this statement was supported only by intuition and lacked approximation guarantees. 

\subparagraph*{Contributions.} 
In this paper, we are interested in developing theoretical understandings between the Reeb space and its discrete approximations to support its use in practical data analysis. 
Using tools from category theory, we formally prove the convergence between the Reeb space and mapper in terms of an interleaving distance between their categorical representations (Theorem~\ref{theorem:mapper-convergence}). 
Given a sequence of refined discretizations, we prove that these approximations converge to the Reeb space in the interleaving distance; this also helps to quantify the approximation quality of the discretization at a fixed resolution.  
Such a result easily generalizes to special cases of mapper such as the JCN. 
Our work extends and generalizes the tools from the categorical representation of Reeb graphs~\cite{SilvaMunchPatel2014} to a new categorical framework for Reeb spaces.  
In particular, we provide for the first time the definition of the interleaving distance for Reeb spaces (Definition~\ref{definition:interleaving}). We demonstrate that such a distance is an extended pseudometric (Theorem~\ref{theorem:ExtendedPseudometric}) and it provides a simple and formal language for structural comparisons. 
Finally in the setting of Reeb graphs (when $d = 1$), we demonstrate that mapper converges to the Reeb graph geometrically on the space level (Corollary~\ref{theorem:geometric-convergence}). We further provide an algorithm for constructing a continuous representation of mapper geometrically from its categorical representation.


\section{Topological Notions}
\label{sec:Topology}

We now review the relevant background on the Reeb space~\cite{EdelsbrunnerHarerPatel2008, Patel2010} and mapper\footnote{Mapper was originally referred to as a method~\cite{SinghMemoliCarlsson2007}, however we refer to it as a topological construction/object in this paper.}~\cite{DeyMemoliWang2015, SinghMemoliCarlsson2007}.   
In theory, we assume the data given is a compact topological space $\X$ with an $\R^d$-valued function, $f:\X \to \R^d$, often denoted $(\X,f)$.
In practice, we assume the data we work with is a multivariate PL mapping $f$ defined over a simplicial mesh; 
more restrictively (for easier exposition of our algorithms and proofs), we consider a generic, PL mapping $f$ from a combinatorial manifold~\cite{Penna1978}
to $\Rspace^d$.

\subparagraph*{Reeb Space.} 
Let  $f: \Xspace \to \Rspace^d$ be a generic, continuous mapping\footnote{For simplicity, assume $f$ is a PL mapping defined on a combinatorial manifold.}. 
Intuitively, the Reeb space of $f$ parametrizes the set of components of preimages of points in $ \Rspace^d$~\cite{EdelsbrunnerHarerPatel2008}. 
Two points $x, y \in \Xspace$ are {equivalent}, denoted by $x \sim_f y$, if $f(x) = f(y)$ and $x$ and $y$ belong to the same path connected component of the preimage, $f^{-1}(f(x)) = f^{-1}(f(y))$.  
The \emph{Reeb space} is the quotient space obtained by identifying equivalent points, 
that is, $\RR(\X,f) = \Xspace/\sim_f$, together with the quotient topology inherited from $\Xspace$. 
A powerful analysis tool, the \emph{Reeb graph}, can be considered a special case in this context when $d = 1$. 
Reeb spaces have been shown to have triangulations and canonical stratifications into manifolds for nice enough starting data~\cite{EdelsbrunnerHarerPatel2008}.  

\subparagraph*{Mapper.} 
An open cover of a topological space $\Xspace$ is a collection $\Ucal = \{U_{\alpha}\}_{\alpha \in A}$ of open sets for some indexing set $A$ such that $\bigcup_{\alpha \in A} U_{\alpha} = \Xspace$. 
In this paper, we will always assume that each $U_{\alpha}$ is path-connected and a cover means a finite open cover. 
We define a finite open cover $\UU$ to be a \emph{good} cover if every finite nonempty intersection of sets in $\UU$ is contractible. 
Given a cover $\Ucal = \{U_{\alpha}\}_{\alpha \in A}$ of $\Xspace$, let $\Nerve(\Ucal)$ denote the simplicial complex that corresponds to the \emph{nerve} of the cover $\Ucal$, 
$\Nerve(\UU) = \left\{ \sigma \subseteq A \mid \bigcap_{\alpha \in \sigma} U_\alpha \neq \emptyset\right\}$. 
Given a (potentially multivariate) continuous map $f: \Xspace \to \Yspace$ where $\Yspace$ is equipped with a cover $\Ucal = \{U_{\alpha}\}_{\alpha \in A}$, we write $f^*(\Ucal)$ as the cover of $\Xspace$ obtained by considering the path connected components of $f^{-1}(U_{\alpha})$ for each $\alpha$. 
Given such a function $f$, its \emph{mapper construction} (or \emph{mapper} for short) $M$ is defined to be the nerve of  $f^*(\Ucal)$, $M(\Ucal, f) : = \Nerve(f^*(\Ucal))$~\cite{SinghMemoliCarlsson2007}. 
Intuitively, considering a real-valued function $f: \Xspace \to \Rspace$ and a cover $\Ucal_{\e}$ of $\image(f) \subseteq \Rspace$ consisting of  intervals of length at most $\e$, the corresponding mapper $M(\Ucal_{\e}, f)$ can be thought of as a relaxed Reeb graph that has been conjectured to converge to the Reeb graph of $f$ as $\e$ tends to zero~\cite{DeyMemoliWang2015, SinghMemoliCarlsson2007}, although no formal proofs have been previously provided. 


\section{Categorical Notions}
\label{sec:categoryTheory}


\subparagraph*{Category and opposite category.} Category theory~\cite{MacLane1978} can be thought of as a generalization of set theory in the sense that the item of study is still a set (technically a proper class),  
but now we are additionally interested in studying the relationships between the elements of the set. 
Mathematically, a \emph{category} is an algebraic structure that consists of mathematical \emph{objects} with a notion of \emph{morphisms} (colloquially referred to as \emph{arrows}) between the objects.
A category has the ability to compose the arrows associatively, and there is an identity arrow for each object. 
Examples are abundant and those important 
\begin{wrapfigure}{r}{2.7in}
\includegraphics[width = 2.7in]{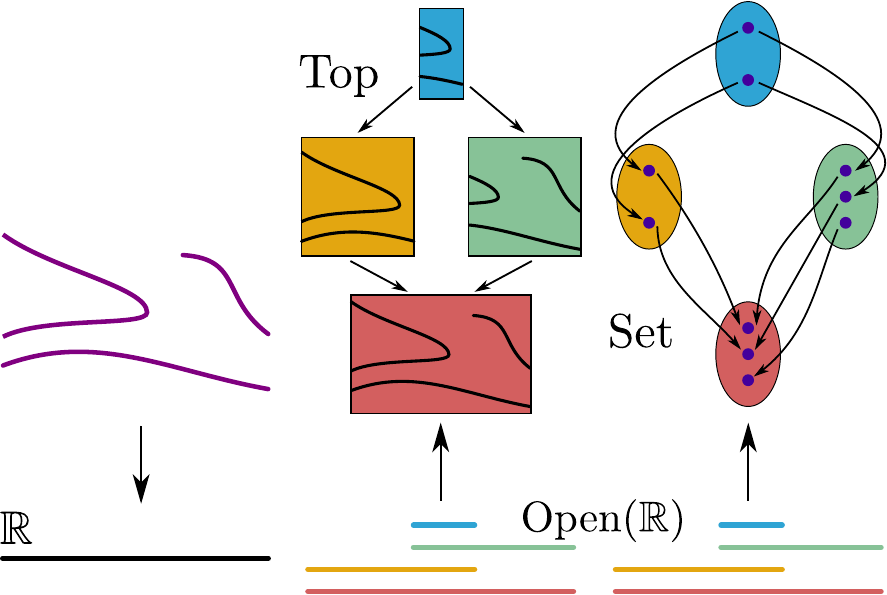}
\caption{
The data of a Reeb graph (on the left) can be stored as a functor.  First, we give the middle functor $f^{-1}: \Open(\R) \to \Top$ which sends each open set $I$ to the topological space $f\inv(I)$; and sends each inclusion map between open sets $I \subseteq J$ to an inclusion map $f\inv(I)\to f\inv(J)$. Then the Reeb graph information is represented by composing this functor with the functor $\pi_0: \Top \to \Set$, 
producing a functor on the right $ \pi_0 f^{-1} : \Open(\R) \to \Set$.  
Via $\pi_0$, the inclusion maps on the topological spaces become set maps.
}
\label{fig:cosheafCondition}
\end{wrapfigure}
to our exposition are: 
the category of topological spaces (as the objects) with continuous functions between them (as the arrows), denoted as $\Top$; 
the category of sets with set maps, denoted as $\Set$; 
the category of open sets in $\R^d$ with inclusion maps, denoted as $\Open(\R^d)$; 
the category of vector spaces with linear maps, denoted as $\Vect$; and the category of real numbers with inequalities connecting them, denoted as $\Rcat$. 
In addition, any simplicial complex $K$ induces a category $\Cell(K)$ where the objects are the simplices of $K$, and there is a morphism $\sigma \to \tau$ if $\sigma$ is a face of $\tau$.
Intuitively, we could think of a category as a big (probably infinite) directed multi-graph with extra underlying structures (due to the associativity and identity axioms obeyed by the arrows):   
the objects are the nodes, and each possible arrow between the nodes is represented as a directed edge. 
One common example used extensively throughout this paper is the idea of a \emph{poset category}, which is a category $\PCat$ in which any pair of elements $x,y \in \PCat$ has at most one arrow $x \to y$.  
Categories such as $\Open(\R^d)$ and  $\Rcat$ are poset categories since there is exactly one arrow $I \to J$ between open sets if $I \subseteq J$ and exactly one arrow $a \to b$ between real numbers if $a \leq b$.
We often abuse notation and denote arrows in this category by the relation providing the poset structure, e.g.~$I \subseteq J$ instead of $I \to J$ and $a \leq b$ instead of $a \to b$.
In the graph description, a poset category can be thought of as a directed graph which is not a multigraph.
The \emph{opposite category } (or dual category) $\CC^{op}$ of a given category $\CC$ is formed by reversing the arrows (morphisms), i.e. interchanging the source and target of each arrow.

\subparagraph*{Functor.} 
A \emph{functor} is a map between categories that maps objects to objects and arrows to arrows. 
A functor $F: \CC \to \DD$ for categories $\CC$ and $\DD$ maps an object $x$ in $\CC$ to an object $F(x)$ in $\DD$, and maps an arrow $f: x \to y$ of $\CC$ to an arrow $F[f]: F(x) \to F(y)$ of $\DD$ in a way that respects the identity and composition laws.
In the above graph allegory, a functor is a map between graphs which sends nodes (objects) to nodes and edges (arrows) to edges in a way that is compatible with the structure of the graphs.
An example of a functor is the homology functor $H_p: \Top \to \Vect$ which sends a topological space $\X$ to its $p$-th singular homology group $H_p(\X)$ (a vector space assuming field coefficients), and  sends any continuous map $f: \X \to \Y$ to the linear map between homology groups, $H_p[f] := f_*:H_p(\X) \to H_p(\Y)$. 
Another functor used extensively in this paper is $\pi_0:\Top \to \Set$ which sends a topological space $\X$ to a set $\pi_0(\X)$ where each element represents a path connected component of $\X$, and  sends a map $f:\X \to \Y$ to a set map $\pi_0[f] := f_*:\pi_0(\X) \to \pi_0(\Y)$.

\begin{wrapfigure}{r}{1.4in}
\begin{center}
\includegraphics{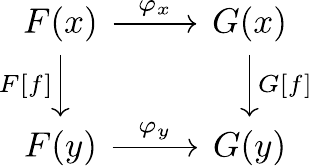}
\end{center}
 \caption{The diagram for a natural transformation.}
 \label{fig:natTrans}
\end{wrapfigure}
\subparagraph*{Natural transformation.}
We can make any collection of functors of the form $F: \CC \to \DD$ into a category by defining arrows between the functors.
A \emph{natural transformation} $\phi: F \Rightarrow G$ between functors $F,G: \CC \to \DD$ is a family of arrows $\phi$ in $\DD$ such that (a) for each object $x$ of  $\CC$, we have $\phi_x: F(x) \to G(x)$, an arrow of $\DD$; and (b) for any arrow $f: x\to y$ in $\CC$, $G[f] \circ \phi_x = \phi_y \circ F[f]$, that is, the diagram of Figure~\ref{fig:natTrans} commutes. 
Any collection of functors $F: \CC \to \DD$ can thus be turned into a category, with the functors themselves as objects and the natural transformations as arrows, 
notated as $\DD^\CC$. 
This notation is used heavily throughout this paper where  always $\DD = \Set$. 
If for every object $x$ of $\CC$, the arrow $\phi_x$ is an isomorphism in $\DD$, then $\phi$ is a \emph{natural isomorphism} (equivalence) of functors.  Two functors $F$ and $G$ are \emph{(naturally) isomorphic} if there exists a natural isomorphism from $F$ to $G$.

\subparagraph*{Categorical Reeb graph.} 
For a real-valued function $f:\X \to \R$, 
the data of its corresponding Reeb graph can be stored as a functor $F:= \pi_0f\inv: \Open(\R) \to \Set$,  defined by sending each open set $I$ to a set $F(I) := \pi_0f\inv(I)$ that contains all the path connected components of $f^{-1}(I)$; and by sending an inclusion  $I \subseteq J$ to a set map $F[I \subseteq J]: F(I) \to F(J) $ induced by the inclusion $f\inv(I) \subseteq f\inv(J)$. 
This is illustrated in Figure~\ref{fig:cosheafCondition}. 
The objects $F(I) $ store the connected components sitting over any open set; 
the information from the arrows $F(I) \to F(J)$ gives the information needed to glue together all of this data. 
This construction produces a categorical representation of the Reeb graph, referred to as the \emph{categorical Reeb graph}. 
It was used in \cite{SilvaMunchPatel2014} to define the interleaving distance for Reeb graphs which we generalize to Reeb spaces in Section \ref{sec:ReebSpace}.

\begin{wrapfigure}{r}{1.7in}
  \vspace{-2mm}
 \begin{center}
 \includegraphics{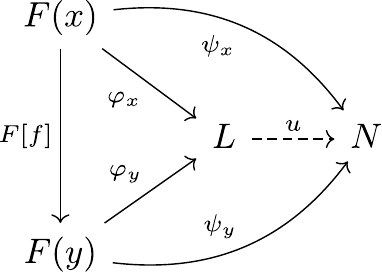}
 \end{center}
 \vspace{-4mm}
 \caption{Defining a colimit.}
 \label{fig:colimit}
 \vspace{-2mm}
 \end{wrapfigure}

\subparagraph*{Colimit.} 
The final category theoretic notion necessary for our results are colimits.
The \emph{cocone} $(N, \psi)$ of a functor $F: \CC \to \DD$ is an object $N$ of $\DD$ along with a family of $\psi$ of arrows $\psi_x: F(x) \to N$ for every object $x$ of $\CC$,  such that for every arrow $f: x \to y$ in $\CC$, we have $\psi_y \circ F[f] = \phi_x$. 
We say that a cocone $(N, \psi)$ factors through another cocone $(L, \phi)$ if there exists an arrow $u: L \to N$ such that  $u \circ \phi_x = \psi_x$ for every $x$ in $\CC$.
The \emph{colimit}  of $F: \CC \to \DD$, denoted as $\colim F$, is a cocone $(L, \phi)$ of $F$ such that for any other cocone $(N, \psi)$ of $F$, there exists a unique arrow $u: L \to N$ such that $(N,\psi)$ factors through $(L,\phi)$.
In other words, the diagram of Figure~\ref{fig:colimit} commutes. 
We often  abuse notation by using $\colim F$ to represent 
just the object $L$.
The colimit is universal; in particular, this means that if the colimit $(L,\phi)$ factors through another cocone $(M,\delta)$, then $L$ is isomorphic to $ M$ and the isomorphism is given by the unique arrow $u':M \to L$ that defines it.
We will use this property in the proof of Lemma \ref{lemma:restatePkCk}.

Because we often wish to consider these colimits over a full subcategory $\mathcal{A} \subseteq \CC$, we will denote the restriction as $\colim_{A \in \mathcal{A}} F(A) $.
The properties of a colimit also imply that if we have nested subcategories $\mathcal{A} \subseteq \mathcal{B}$ ($\subseteq \CC$), then there is a unique map $\colim_{A \in \mathcal{A}} F(A) \to \colim_{B \in \mathcal{B}} F(B) $ since we can consider $\colim_{B \in \mathcal{B}} F(B)$ as cocone over $\mathcal{A}$.


\section{Main Results Overview}
\label{sec:Overview}

\begin{wrapfigure}{r}{3.5in}
\vspace{-.4in}
\begin{center}
\includegraphics{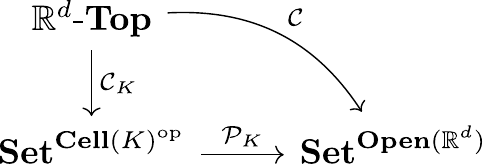}

\vspace{2mm}

\includegraphics{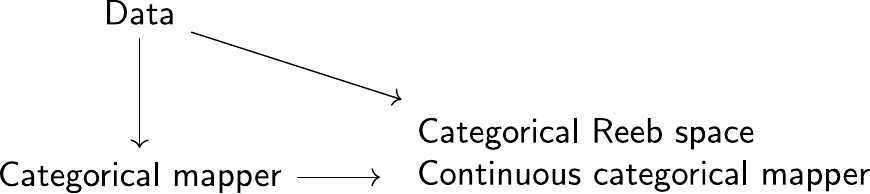}
 \end{center}
 
 \vspace{-5mm} 
 \caption{The diagram for connecting categorical representations of the Reeb space and the mapper. Note that the diagram is \emph{not} commutative.  Theorem \ref{theorem:mapper-convergence} measures the amount that this diagram deviates from being commutative.}
 \label{fig:roadmap}
\vspace{-2mm}
 \end{wrapfigure}
 
The main focus of this paper is to provide a convergence result between the continuous Reeb space and the discrete mapper. 
We define their distance as the interleaving distance between their corresponding categorical representations and emphasize that neither the Reeb space nor the interleaving distance must ever be computed for this result. 
Instead, we provide a theoretical bound on the distance which requires only knowledge of the resolution of  the cover. 
To define the desired distance measure, we use the diagram in Figure~\ref{fig:roadmap} as our roadmap. The remainder of this section is dedicated to describing the various categories at the nodes of the diagram as well as the functors that connect them.

\subparagraph*{Data.}
In our context, data comes in the form of a topological space $\Xspace$ with an $\R^d$-valued  mapping, called an $\R^d$-space.  
We store such data in the category $\Rdtop$. 
Specifically, an object of $\Rdtop$ is a pair consisting of a topological space $\X$ with a continuous map $f:\X \to \R^d$, denoted as $(\X,f)$.
An arrow in $\Rdtop$,  $\nu:(\X,f) \to (\Y,g)$, is a function-preserving map; that is, it is a continuous map on the underlying spaces $\nu:\X \to \Y$ such that $g \circ \nu(x) = f(x)$ for all $x \in \Xspace$.  
Note that many nice constructions such as PL functions on simplicial complexes or Morse functions on manifolds are objects in $\Rdtop$.

\subparagraph*{Categorical Reeb space and its contruction.}
Recall the categorical representation of a Reeb graph is a functor $\Open(\R) \to \Set$. 
In order to define a categorical representation of the Reeb space, we need a higher dimensional analogue of $\Open(\R)$, namely, $\Open(\R^d)$. 
$\Open(\R^d)$ is a category with open sets $I \subseteq \Rspace^d$ as objects, and a unique arrow $I \to J$ if and only if $I \subseteq J$; that is, $\Open(\R^d)$ is a poset category.  
The data of the Reeb space can be stored as a functor $\pi_0f\inv: \Open(\R^d) \to \Set$, 
defined by sending each open set $I$ to a set $\pi_0f\inv(I)$ representing the path connected components of $f^{-1}(I)$; 
and by sending the inclusion arrow $I \subseteq J$ to a set map $\pi_0f\inv(I) \to \pi_0f\inv(J)$ induced by the inclusion $f^{-1}(I) \subseteq f^{-1}(J)$.  
These functors, referred to as the \emph{categorical Reeb spaces}, become objects of the category of functors $\Set^{\Open(\R^d)}$. 

Constructing a Reeb space from the data is now represented by the functor $\CC: \Rdtop \to \Set^{\Open(\R^d)}$ in Figure~\ref{fig:roadmap}. 
In particular, $\CC$ maps an object $(\Xspace, f)$ in $\Rdtop$, representing the data, to a functor 
$F: \Open(\R^d) \to \Set$ in $\Set^{\Open(\R^d)}$, representing its corresponding Reeb space. 
The functor $\CC$ restricts to the Reeb graph construction when $d=1$ \cite{SilvaMunchPatel2014}.
In addition, from the generalized persistence module framework \cite{BubenikdeSilvaScott2014}, we can also extend the idea of the interleaving distance between Reeb graphs (in the case $d = 1$) to these categorical Reeb spaces (in the case $d\geq 1$). 
The definition of functor $\CC$ and the Reeb space interleaving distance are covered in Section \ref{sec:ReebSpace}.

\begin{figure}
 \begin{center}
  \includegraphics[width = \textwidth]{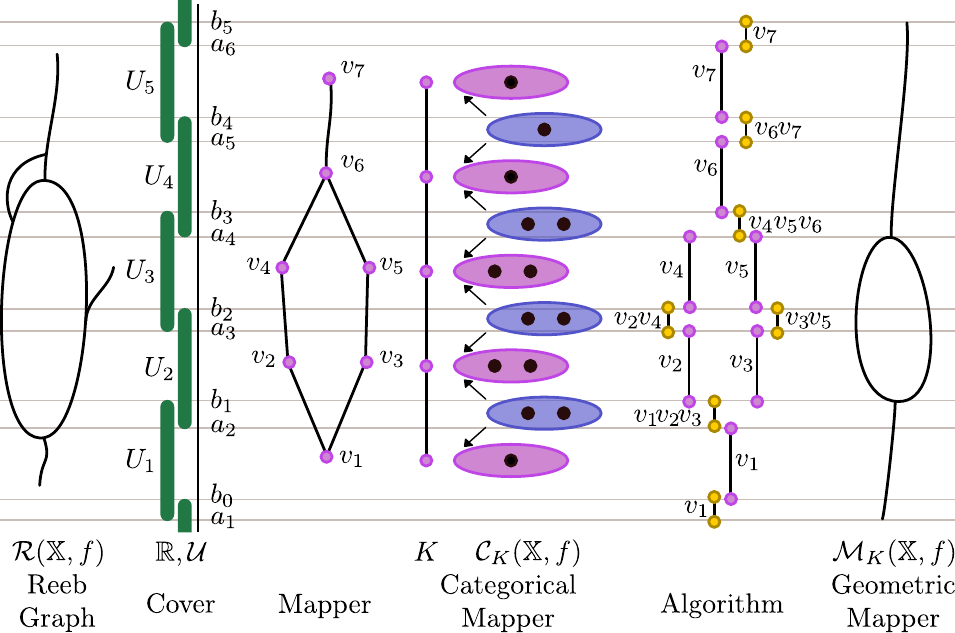}
 \end{center}
 \caption{An example of a Reeb space for $d=1$ (a Reeb graph), denoted as $\RR(\Xspace, f)$, is shown on the left. 
Its associated data $(\Xspace, f)$ is an object in $\Rdtop$ with function $f$ given by height. 
A cover $\UU$ is shown by the green intervals, and the corresponding mapper is shown to its right. The mapper data is equivalently stored as the $C_K(\Xspace,f)$ functor defined on an abstract simplicial complex $K = \mathrm{Nrv}(\UU)$. 
Note that although we draw $K$ in the same plane as the other objects, it does not have a geometric embedding, nor does it have a natural map to $\R$.
This is remedied with the geometric representation of this data, $\MM_K(\Xspace,f):= \DD \PP_K \CC_K (\Xspace, f)$ which is shown at the far right. Corollary~\ref{theorem:geometric-convergence} asserts that the interleaving distance between the leftmost and rightmost graphs is bounded by $\e = \res(\UU)$.}
 \label{fig:ReebGraph}
\end{figure}

\subparagraph*{Categorical mapper and its construction.}
Instead of working with continuous objects, we can instead choose a discretization  represented by a simplicial complex $K$.  
Given a cover $\UU = \{U_\alpha\}_{\alpha \in A}$ for $\image(f) \subseteq \R^d$, let $K = \Nerve(\UU)$. Through the machinery detailed in Section \ref{sec:categorical-mapper},
we create a categorical representation of the mapper (referred to as the \emph{categorical mapper}) as a functor $F: \Cell(K)^\op \to \Set$ (an object of $\Set^{\Cell(K)^\op}$); 
and such a construction is represented by the $\CC_K$ functor\footnote{A related but slightly different categorical mapper was introduced by Stovner~\cite{Stovner2012}, as a functor from the category of covered topological spaces to the category of simplicial complexes.}. 

\subparagraph*{Comparing Reeb space and mapper.}
It should be noted that the Reeb space and  the mapper are inherently different objects.  
The Reeb space comes equipped with an $\R^d$-valued function, while there is no such function built into the mapper even though its construction is highly dependent on the functions chosen to partition the data set~\cite{SinghMemoliCarlsson2007}. 
In particular, the two objects are in completely different categories.
So, to compare these objects, we study the image of the categorical mapper under the functor $\PP_K$, which turns the categorical mapper (a discrete object) into a continuous one comparable with the categorical Reeb space. 
In particular, for data given as $(\Xspace, f)$ in $\Rdtop$, we compare its image in $\Set^{\Open(\Rspace^d)}$ via the functor $\CC$, to its image in $\Set^{\Open(\Rspace^d)}$ via the functor $\PP_K\CC_K$. 
Symbolically, following Figure~\ref{fig:roadmap}, we are comparing $\PP_K\CC_K(\X,f)$ to $\CC(\X,f)$. 
This relationship and the construction of functor $\PP_K$ are covered in Section \ref{sec:convergence}. 

We then prove our main result,  the categorical convergence theorem below. 

\begin{theorem}[Convergence between Categorical Reeb Space and Categorical Mapper]
\label{theorem:mapper-convergence}
Given a multivariate function $f: \X \to \R^d$ defined on a compact topological space\footnote{For simplicity, we assume a combinatorial $s$-manifold; however this is not necessary for the proof.}, the data is represented as an object $(\X, f)$ in $\Rdtop$.
Let $\UU = \{ U_{\alpha}\}_{\alpha \in A}$ be a good cover of $f(\X) \subseteq \R^d$, $K$ be the nerve of the cover and $\mathrm{res}(\UU)$ be the resolution of the cover, that is, the maximum diameter of the sets in the cover $ \res(\UU) = \sup \{ \diam(U_\alpha) \mid U_\alpha \in \UU\}$. 
Then 
$$d_I(\CC(\X, f), \PP_K\CC_K(\X, f)) \leq \mathrm{res}(\UU).$$
\end{theorem}

Theorem~\ref{theorem:mapper-convergence} states that for increasingly refined covers, the image of the categorical mapper converges to the categorical Reeb space in the interleaving distance. 
In other words, the distance between the mapper and the Reeb space is bounded above by the resolution of the discretization.  Thus, we can make approximation guarantees about the accuracy of the mapper based on a property of the chosen discretization.

\subparagraph*{Summary.}
The various categorical representations can be summarized in Figure~\ref{fig:roadmap}, some of which are illustrated in Figure~\ref{fig:ReebGraph} for the case when $d = 1$. 
The initial data received is an object $(\Xspace, f)$ in $\Rdtop$. 
 Then we can either construct its categorical Reeb space through the functor $\CC$, or construct its categorical mapper using the functor $\CC_K$.
 In order to compare these two objects in the same category, we push the mapper along using the $\PP_K$ functor, and then compute the distance between $\CC(\Xspace, f)$ and $\PP_K \CC_K (\Xspace, f)$ in $\Set^{\Open(\Rspace^d)}$.
We should stress before we continue that the diagram of Figure \ref{fig:roadmap} does not commute.
 In a way, the above distance is measuring how far  the diagram is from being commutative. 
Making no assumptions about $\UU$, Theorem~\ref{theorem:mapper-convergence} states that the interleaving distance between the results of the two paths in the diagram is bounded by the resolution of $\UU$. 
Furthermore in Section \ref{sec:GeometricReps}, for the special case when $d = 1$, we turn our categorical convergence theorem,  Theorem~\ref{theorem:mapper-convergence}, into the geometric convergence theorem, Corollary~\ref{theorem:geometric-convergence}. 
Finally, we provide an algorithm for producing a geometric representation of the image of categorical mapper, $\PP_K\CC_K(\X,f)$.


\section{Interleaving Distance between Reeb Spaces}
\label{sec:ReebSpace}

As described in Section \ref{sec:Overview}, we start by generalizing the categorical Reeb graph to the categorical Reeb space. 
Given the data received as a topological space $\X$ equipped with an $\Rspace^d$-valued function $f: \X \to \R^d$, denoted as $(\X, f)$, we define the functor $\CC: \Rdtop \to \Set^{\Open(\R^d)}$ as follows: 
$\CC$ maps an object $(\Xspace, f)$ in $\Rdtop$ to a functor $\CC(\Xspace, f) : = \pi_0f\inv: \Open(\R^d) \to \Set$ in $\Set^{\Open(\R^d)}$, 
and an arrow $\nu: (\Xspace, f) \to (\Yspace, g)$ to a natural transformation $\CC[\nu]$ induced by the inclusion $\nu f\inv(I) \subseteq g\inv(I)$.
The functor $\CC$ turns the given data into the categorical representation of the Reeb space, and the functoriality of $\pi_0$ makes it a well-defined functor. 

Our first goal is to define the interleaving distance for these categorical Reeb spaces.
Denote the $\e$-thickening of a open set $I \in \Open(\R^d)$ to be the set $I^\e := \{x \in \R^d \mid \|x-I\| < \e\}$.
Using this, we can define a thickening functor $T_\e: \Open(\R^d) \to \Open(\R^d)$ by $T_\e(I) := I^\e$, and $T_\e[I \subseteq J] := \{ I^\e \subseteq J^\e\}$.
Let $\SS_\e$ be the functor from  $\Set^{\Open(\R^d)} $ to itself defined by $\SS_\e(\FF) := \FF T_\e$, for every functor $\FF: \Open(\R^d) \to \Set$. 
Given the two functors $\FF$ and $ \SS_{2\e}(\FF)$, both of which are defined on $\Open(\R^d) \to \Set$, 
there is an obvious natural transformation $\eta:\FF \Rightarrow \SS_{2\e}\FF$ defined by $\eta_I = \FF[I \subseteq I^{2\e}]$. 
We write $\tau: \GG \Rightarrow \SS_{2\e}(\GG)$ for the analogous natural transformation for $\GG$.

\begin{definition}[Interleaving distance between Categorical Reeb spaces]
\label{definition:interleaving}
 An $\e$-interleaving between functors $\FF,\GG: \Open(\R^d) \to \Set$ is a pair of natural transformations, 
$\phi:\FF \Rightarrow \SS_\e(\GG)$ and 
$\psi:\GG \Rightarrow \SS_\e(\FF)$
such that the diagrams below commute.  
\begin{center}
\includegraphics{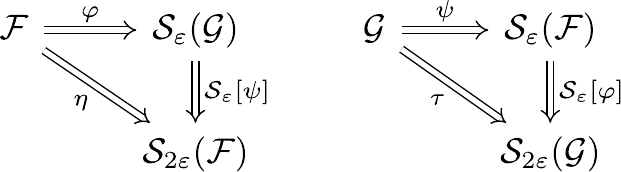}
\end{center}
\noindent Given two functors  $\FF,\GG:\Open(\R^d) \to \Set$, the interleaving distance is defined to be 
$$  d_I(\FF,\GG) =\inf\{ {\e \in \R_{\geq0}} \mid \FF,\GG\textrm{ are }\e\textrm{-interleaved}\}.$$
\noindent We define $d_I(F,G) = \infty$ if the set on the right-hand side is empty. 
\end{definition}

We prove in the full version~\cite{MunchWang2016} the following property of $d_I$ using \cite{BubenikdeSilvaScott2014}. 
\begin{theorem}
The interleaving distance $d_I$, between two categorical representations of Reeb spaces, is an extended pseudometric on $\Set^{\Open(\R^d)}$.
\label{theorem:ExtendedPseudometric}
\end{theorem}

\subparagraph*{Special case for Reeb graphs.}
When $d=1$ we have much more control of the situation.  
In particular,  \cite{SilvaMunchPatel2014} gives us that the category of Reeb graphs, defined to be finite graphs with real valued functions that are strictly monotone on the edges, is equivalent to a well-behaved subcategory of  $\Set^{\Open(\R)} $.   
Theorem~\ref{theorem:interleavingDistance} (as a direct consequence of Corollary 4.9 in~\cite{SilvaMunchPatel2014}) says that the above defined interleaving distance $d_I$ is an extended metric, not just a pseudometric, when restricted to these objects.

\begin{theorem}[\cite{SilvaMunchPatel2014}]
\label{theorem:interleavingDistance}
When $d=1$, $d_I(\CC(\X,f),\CC(\Y,g))$ is an extended metric on the categorical Reeb spaces.  
\end{theorem}

Theorem~\ref{theorem:interleavingDistance} means that for $d = 1$,  if $d_I(\CC(\X,f), \PP_K\CC_K(\X,f)) = 0$ (that is, when the categorical mapper converges to the categorical Reeb graph), then $\CC(\X,f)$ and $\PP_K\CC_K(\X,f)$ are isomorphic as functors. 
This implies that, in the special case when $d = 1$,  mapper converges to the Reeb graph as spaces, not just in the interleaving distance. 
While recent work is beginning to elucidate the case where $d>1$, the technical finesse needed to make a similar statement to Theorem~\ref{theorem:interleavingDistance} is beyond the scope of this paper.  
Thus, we will stick to statements about the categorical representations for Reeb spaces when $d>1$, and make concrete geometric statements when they are available for $d=1$ (see Section \ref{sec:GeometricReps}).


\section{Categorical Representation of Mapper and its Construction}
\label{sec:categorical-mapper}

The beauty of working with category theory is that we can store a categorical representation of the mapper as sets over the nerve of a cover, rather than working directly with its complicated topological definition (given in Section \ref{sec:Topology}).
Given a choice of finite open cover for $\image(f) \subseteq \Rspace^d$, $\UU = \{U_\alpha\}_{\alpha \in A}$, let $K = \Nrv(\UU)$.
In order to ensure that $K$ faithfully represents the underlying structure, we will assume that $\UU$ is a \emph{good} cover. 
This ensures that the nerve lemma applies; that is, $K$ has the homotopy type of $\image(f) \subseteq \Rspace^d$ (see, e.g., Corollary 4G.3~\cite{Hatcher2002} or Theorem 15.21~\cite{Kozlov2008}). 

For simplicity of notation, we denote $\UU_\sigma = \bigcap_{\alpha \in \sigma} U_{\alpha}$ to be the open set in $\R^d$ associated to the simplex $\sigma \in K$.
One important property of this construction is that for $\sigma \leq \tau$ in $K$, the associated inclusion of spaces is reversed: $\UU_\sigma \supseteq \UU_\tau$.
So, if we wish to represent the connected components for a particular $\UU_\sigma$ for $\sigma \in K$, we can still consider $\pi_0f\inv(\UU_\sigma)$, however, the face relation $\sigma \leq \tau$ induces a ``backwards'' mapping $\pi_0f\inv(\UU_\tau) \to \pi_0f\inv(\UU_\sigma)$.
We keep track of this switch using the opposite category.
Recall $\Cell(K)$ is a category with simplices of $K$ as objects and a unique arrow $\sigma \to \tau$ given by the face relation $\sigma \leq \tau$.
Then the opposite category, $\Cell(K)^{\op}$, has the simplices of $K$ as objects and a unique arrow $\tau \to \sigma$ given by the face relation $\sigma \leq \tau$.

Thus, given  an object $(\Xspace, f)$ in $\Rdtop$, we have a functor $\CC_K^f: \Cell(K)^\op \to \Set$ that maps every $\sigma$ to $\CC_K^f(\sigma) : = \pi_0f\inv(\UU_\sigma)$. 
We are required to use the opposite cell category so that $\CC_K^f$ maps the morphism $\sigma \leq \tau$ (equivalently notated $\tau \to \sigma$ in the opposite category) to the set map 
 $\pi_0f\inv(\UU_\tau) \to \pi_0f\inv(\UU_\sigma)$
induced by the inclusion $\UU_\tau \subseteq \UU_\sigma$ as discussed above.
This functor is used to represent  the categorical mapper of $(\X,f)$ for the cover $\UU$.

Note that the functor $\CC_K^f$ is an object of  the category of functors $\Set^{\Cell(K)^\op}$. 
The process of building the mapper is thus represented itself by the functor $\CC_K: \Rdtop \to \Set^{\Cell(K)^\op}$, which is defined as follows.
For the objects, $\CC_K$ maps an $\R^d$-space $(\Xspace, f)$ in $\Rdtop$ to the functor $\CC_K(\X, f) := \CC_K^f$ as given above. 
For the morphisms, it sends a function preserving map $\nu: (\X,f) \to (\Y,g)$ to a natural transformation (which is an arrow in $\Set^{\Cell(K)^\op}$), $\CC_K[\nu]:\CC_K^f\to \CC_K^ g$. 
Technical details in checking that $\CC_K[\nu]$ is indeed a natural transformation are deferred to the full version~\cite{MunchWang2016}. 


\section{Convergence between Mapper and Reeb Space}
\label{sec:convergence}

In order to compare the discrete mapper with the continuous Reeb space, we must move them both into the same category.
At the moment, for data given as $(\Xspace, f)$ in $\Rdtop$, we have the categorical Reeb space representation $\CC(\Xspace, f)$ in $\Set^{\Open(\R^d)}$, and the categorical mapper representation $\CC_K(\Xspace, f)$ in $\Set^{\Cell(K)^\op}$.
Thus we must first define the functor $\PP_K$ in order to push the mapper representation into the $\Set^{\Open(\R^d)}$ category, then prove the convergence result there using the interleaving distance from Section \ref{sec:ReebSpace}.
Here, we will give the definition of $\PP_K$ coming from the categorical setting, and then give an equivalent functor $\FF$ in Lemma \ref{lemma:restatePkCk} which is more intuitive to work with.

Given an abstract simplicial complex $K$ which is the nerve of the cover $\UU$, we define $K_{A}$ for a open set $A \subseteq \R^d$ to be the collection of simplices in $K$ such that the associated intersection $\UU_\sigma$ intersects $A$, $K_A = \{\sigma \in K \mid \UU_\sigma \cap A \neq \emptyset\}$ (see the full version~\cite{MunchWang2016} for an example when $d=2$). 
Now we can construct the functor $\PP_K: \Set^{\Cell(K)^\op} \to \Set^{\Open(\R^d)}$ as follows.
Given a functor $F: \Cell(K)^\op \to \Set$,  
$\PP_K$ sends it to a functor $\PP_K(F): \Open(\R^d) \to \Set$ by defining 
$$\PP_K(F)(I) = \colim_{\sigma \in K_I} F(\sigma)$$ for every $I$ in $\Open(\R^d)$. 
Here, the colimit construction can be thought of as a set representing the connected components over the collection of open sets $\UU_\sigma$ for the simplices $\sigma \in K_I$, or equivalently, over the union $\bigcup_{\sigma \in K_I}{\UU_\sigma}$.
The morphisms in the two functor categories $\Set^{\Cell(K)^\op}$ and $\Set^{\Open(\R^d)}$ are natural transformations; $\PP_K$ sends arrows to arrows in a well-defined way via the colimit as discussed at the end of Section \ref{sec:categoryTheory}, since if $I \subseteq J$, then $K_I  \subseteq K_J$. 
Additionally, we must check that $\PP_K$ sends a natural transformation $\eta:F \Rightarrow G$ to a natural transformation $\PP_K(F) \to \PP_K(G)$; we omit this bookkeeping here. 
Since  mapper depends on the choice of a cover, it makes sense that the cover and, in particular, its resolution will be a key factor in understanding the convergence.
With all of this machinery, we have our main result, Theorem \ref{theorem:mapper-convergence}. 

Theorem \ref{theorem:mapper-convergence} implies that if we have a sequence of covers $\UU_i$ such that $\mathrm{res}(\UU_i) \to 0$, then the categorical representations of the associated mappers converge to the Reeb space in the interleaving distance. 
Its proof relies on a main technical result,  Lemma~\ref{lemma:restatePkCk} below, which relates the functor $\PP_K\CC_K(\X, f)$ to  one which avoids the combinatorial structure of $K$ as much as possible and instead works with inverse images of subsets of $\R^d$.  

\begin{lemma}
\label{lemma:restatePkCk}
Let $\FF: \Open(\R^d) \to \Set$ be a functor which maps an open set $I$ to a set $\pi_0f\inv(\bigcup_{\sigma \in K_I}\UU_\sigma)$ with 
morphisms induced by $\pi_0$ on the inclusions.
Then, the functor $\PP_K\CC_K(\X, f)$ is equivalent to $\FF$.
\end{lemma} 

\noindent{\color{darkgray}\sffamily\bfseries Proof.}
The functor $\CC_K(\X, f)= \CC_K^f: \Cell(K)^\op \to \Set$ is given by sending a cell $\sigma$ to $\pi_0f\inv(\UU_\sigma)$, and its composition with $\PP_K$ is given by 
$\PP_K\CC_K(\X, f) = \PP_K(\CC_K^f): \Open(\R^d) \rightarrow \Set$ defined by 
$\PP_K(\CC_K^f)(I) =  \colim_{\sigma \in K_I } \CC_K^f(\sigma)$. 
To establish a natural equivalence of functors, 
we will construct a natural transformation $\psi: \FF \Rightarrow \PP_K\CC_K(\X, f)$ which is an isomorphism for each $\psi_I$. 
As a roadmap, we can refer to the following diagram:
 \begin{center}
 \includegraphics{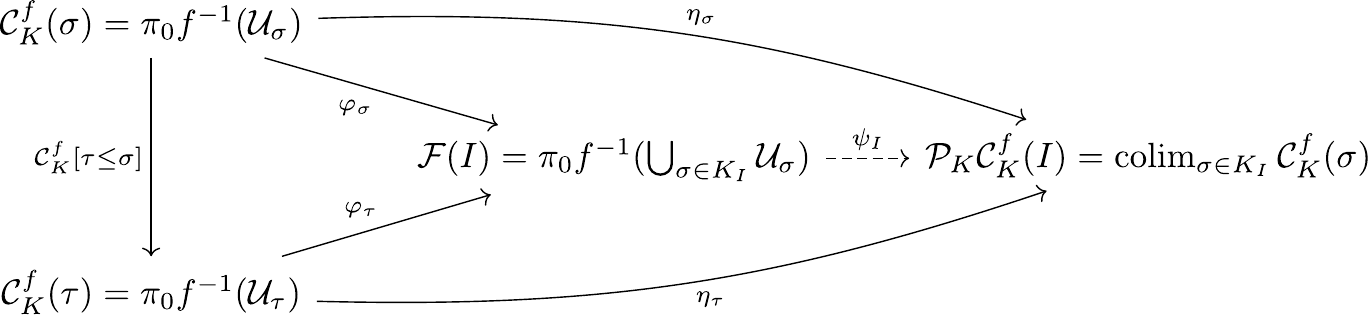}
 \end{center} 
By definition of $\FF$, $\FF(I) = \pi_0f\inv(\bigcup_{\sigma \in K_I}\UU_\sigma)$ so there are obvious maps induced by inclusions $\phi_\sigma: \pi_0f\inv(\UU_\sigma) \to \FF(I)$ which all commute; this gives us a cone $(\FF(I), \phi_\sigma)$ for the diagram $\{\CC_K^f(\sigma)\}_{\sigma \in K_I}$. 
The colimit  of this same diagram is a cocone denoted by $(\PP_K\CC_K^f(I), \eta_{\sigma})$. 
We will construct a map 
$\psi_I: \pi_0f\inv (\bigcup_{\sigma \in K_I} \UU_\sigma) \to \colim_{\sigma \in K_I } \CC_K^f(\sigma)$ 
such that the colimit cocone factors through the cocone $(\FF(I),\phi_\sigma)$ using $\psi_I$; 
that is, $\psi_I \circ \phi_\sigma = \eta_\sigma$ for all $\sigma \in K_I$. 
The universality of the colimit then implies that $\psi_I$ is an isomorphism.

To construct  $\psi_I$, consider any $u$ in $\pi_0f\inv(\bigcup_{\sigma \in K_I} \UU_\sigma)$.  
This set element represents a connected component in $f\inv(\bigcup_{\sigma \in K_I}\UU_\sigma)$, and thus there is at least one $\sigma$ with an element $v \in \CC_K^f(\sigma)$ such that $\phi_\sigma(v) = u$. 
Now we  define $ \psi_I(u) = \eta_\sigma (v)$.
Ensuring that $\psi_I$ above is well defined corresponds to ensuring that if there are $v \in \CC_K^f(\sigma)$ and $v' \in \CC_K^f(\sigma')$ with $\phi_\sigma(v) = \phi_\sigma(v') = u$, then $\eta_\sigma(v) = \eta_{\sigma'}(v')$.  
Note that $v$ and $v'$ represent (path) connected components in $f\inv(\UU_\sigma)$ and $f\inv(\UU_{\sigma'})$ respectively.  
Let $x$ and $x'$ be points in these respective connected components. 
Since these points are in the same connected component of $f\inv(\bigcup_{\sigma \in K_I} \UU_\sigma)$, there is a path connecting them, and thus a finite sequence of $\tau_i \in K_I$ with  $\tau_0 = \sigma$ and $\tau_n = \sigma'$,  such that $f\inv(\UU_{\tau_i})$ covers the path.  
We can additionally assume that the $\tau_i$ give the maximal simplex containing the path at each location, so that $\tau_i \leq \tau_{i+1}$ or $\tau_i \geq \tau_{i+1}$ for each $i$. 
Let $v_i \in \pi_0f\inv\UU_{\tau_i}$ represent the connected component of the path.  
Then we must have $\CC_K^f[\tau_i \leq \tau_{i+1}] (v_{i+1})  = v_i$ or $\CC_K^f[\tau_{i+1} \leq \tau_{i}] (v_{i})  = v_{i+1}$ for each $i$.
By the colimit properties, this implies that $\eta_{\tau_i}(v_i) = \eta_{\tau_j}(v_j)$ for all $i$ and $j$, and thus that $\eta_\sigma(v) = \eta_{\sigma'}(v')$ as desired.

 \begin{wrapfigure}{r}{0.5\textwidth}
 \vspace{-7mm}
\begin{center}
\includegraphics{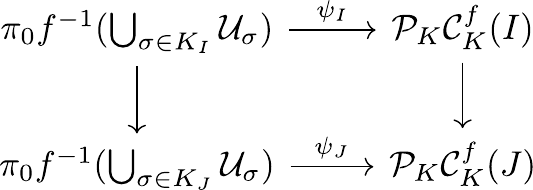}
\end{center} 
\vspace{-6mm}
\caption{The diagram showing that $\phi = \{\phi_I\}$ defines a natural transformation.}
\label{fig:lemma-proof-diagrams}
\vspace{-2mm}
\end{wrapfigure}
 
Finally, we prove that the collection $\{\psi_I\}$ defines a natural transformation. 
Since if $I \subseteq J$, then $K_I \subseteq  K_J$.
Then an exercise in colimit properties ensures that the diagram in Figure~\ref{fig:lemma-proof-diagrams} commutes,
where the arrow on the left is the map induced by inclusions, and the map on the right is induced by the colimit definition.  
\qed

\begin{wrapfigure}{r}{0.56\textwidth}
 \centering
 \includegraphics{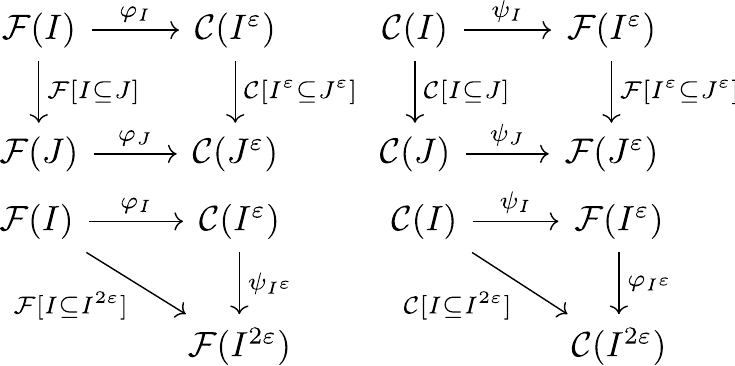}
\caption{Communicative diagrams showing $\phi$ and $\psi$ being natural transformations and $\e$-interleaved.}
\label{fig:proof-diagrams}
\end{wrapfigure}

\noindent{\color{darkgray}\sffamily\bfseries Proof of Theorem \ref{theorem:mapper-convergence}.}
Let $\e = \mathrm{res}(\UU)$. 
Combined with Lemma \ref{lemma:restatePkCk}, 
we will construct,  $\phi: \FF \Rightarrow \CC(\X, f) \circ T_\e$ and $\psi: \CC(\X, f) \Rightarrow \FF \circ T_\e$, and show that they constitute an $\e$-interleaving by showing the diagrams of Figure \ref{fig:proof-diagrams} commute following Definition \ref{definition:interleaving}.

First, we prove the following statement: if $\UU_\sigma \cap I \neq \emptyset$, then $\UU_\sigma \subset I^\e$.
Indeed, for any $x \in \UU_\sigma$, if $x \in I$ then $x \in I^\e$. 
If $x \not \in I$, then because there exists a $y \in \UU_\sigma \cap I$, such that $\|x-y\| \leq \mathrm{diam}(\UU_\sigma) \leq \mathrm{res}(\UU) = \e$, so $x \in I^\e$.
This statement implies that we have the inclusion $\bigcup_{\sigma \in K_I} \UU_\sigma  \hookrightarrow I^\e$.
We define
$\phi_I: \pi_0f\inv\left(  \bigcup_{\sigma \in K_I} \UU_\sigma \right)  \to \pi_0f\inv(I^\e).$

We also have inclusions  
$I \cap f(\X) \hookrightarrow \bigcup_{\sigma \in K_{I}} \UU_\sigma  \hookrightarrow \bigcup_{\sigma \in K_{I^\e}} \UU_\sigma$, 
since any point $x \in I\cap f(\X)$ is contained in some $\UU_\alpha$, for some vertex $\alpha \in K_I \subseteq K_{I^\e}$.
Additionally, since $f\inv(I) = f\inv(I \cap f(\X))$, 
we define 
$\psi_I: \pi_0f\inv(I) \to  \pi_0f\inv\left(  \bigcup_{\sigma \in K_{I^\e}} \UU_\sigma \right)$
to be the composition of the ismorphism $\pi_0f\inv(I) \cong \pi_0f\inv(I \cap f(\X))$ and the map induced by the inclusion $ I \cap f(\X) \hookrightarrow \bigcup_{\sigma \in K_{I^\e}} \UU_\sigma$.

The top left square of Figure~\ref{fig:proof-diagrams} comes from applying the functor $\pi_0f\inv$ to the inclusions
$\bigcup_{\sigma \in K_I}\UU_\sigma \subseteq I^\e \subseteq J^\e$ 
and
$\bigcup_{\sigma \in K_I}\UU_\sigma \subseteq \bigcup_{\sigma \in K_J}\UU_\sigma \subseteq J^\e$ 
for any $I \subseteq J$.
Applying $\pi_0f\inv$ to the inclusions 
$I \cap f(\X) \subseteq \bigcup_{\sigma \in K_{I^\e}} \UU_\sigma \subseteq \bigcup_{\sigma \in K_{J^\e}} \UU_\sigma $ and 
$I \cap f(\X) \subseteq J \cap f(\X) \subseteq \bigcup_{\sigma \in K_{J^\e}} \UU_\sigma $ for $I \subseteq J$, then replacing $\pi_0f\inv(I\cap f(\X))$ and $\pi_0f\inv(J\cap f(\X))$ with the isomorphic $\CC(I)$ and $\CC(J)$ respectively gives the diagram of the top right.
A similar argument implies that 
the diagrams in Figure~\ref{fig:proof-diagrams} bottom also commute, hence $\phi$ and $\psi$ are an $\e$-interleaving.
\qed


\section{Geometric Representations}
\label{sec:GeometricReps}

We now leverage the results of \cite{SilvaMunchPatel2014} to make geometric statements connecting the mapper and the Reeb space for $d = 1$. 
The main idea is to define a mapping that recovers the geometric representation of the mapper from its categorical representation, and to establish convergence between the mapper and the Reeb graph geometrically. Such a mapping relies on well behaved data, made precise by the notion of constructibility. 

\begin{wrapfigure}{r}{3.0in}
\begin{center}
\includegraphics{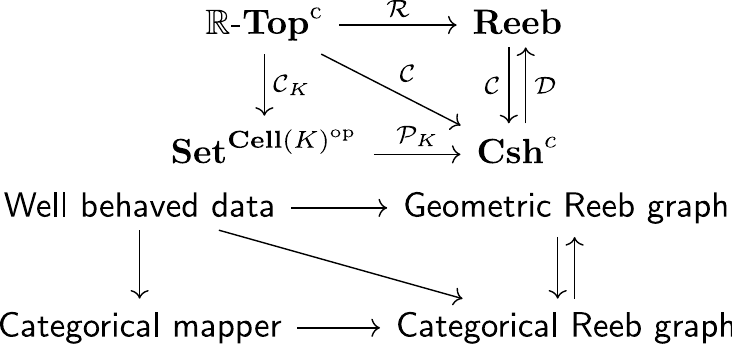}
  \end{center}
 \caption{The diagram for connecting geometric representations of the Reeb graph and the mapper.}
 \label{fig:geometric-roadmap}
 \end{wrapfigure}

\subparagraph*{Review of prior results.} We will follow the notations of \cite{SilvaMunchPatel2014} which occasionally can be technical. 
The categories and functors we will discuss can be summed up in the roadmap of Figure~\ref{fig:geometric-roadmap}. Notice its lower left triangle resembles that of Figure~\ref{fig:roadmap} with further restrictions. 
Recall the notation from Section~\ref{sec:Overview};  when $d=1$, the category $\Rtop$ is exactly the category $\Rdtop$: an object of $\Rtop$ is an $\R$-space (a pair of a topological space $\X$ and a continuous map $f:\X \to \R$), and an arrow in $\Rtop$ is a function-preserving map.

Since the geometric Reeb graph of a general $\R$-space may be
badly behaved, we restrict to special classes of spaces~\cite{SilvaMunchPatel2014}, that is, we focus on well behaved subcategories. 
In particular, we define the full subcategory $\Rtopc$ of $\Rtop$ where the objects are \emph{constructible} $\R$-spaces (see Section 2.2 and Figure 5 of~\cite{SilvaMunchPatel2014} for illustrations and technical details).
This collection includes, e.g., PL functions on triangulations of manifolds and Morse functions.
Then we define the full subcategory $\Reeb$ of $\Rtopc$ (in the finite, discrete setting),  
which is exactly the category of Reeb graphs, viewed as a  graph with a real valued function which is monotone on edges, with arrows given by function preserving maps.
Subsequently, the construction of a (geometric) Reeb graph from well behaved data (a constructible $\R$-space) is captured by the functor $\RR:\Rtopc \to \Reeb$.

We can similarly restrict our objects of interest in $\Set^{\Open(\R)}$ to be well behaved.  
A \emph{cosheaf} is a functor $F: \Open(\R) \to \Set$ such that for any open cover $\UU$ of a set $U$, the unique map $\colim_{U_\alpha \in \UU} F(U_\alpha) \to F(U)$ is an isomorphism.
We further restrict the cosheaves to constructible cosheaves; a cosheaf is \emph{constructible} if there is a finite set $S \subset \R$ such that if $A,B \in \Open(\R)$ with $A \subseteq B$ and $S\cap A = S\cap B$, then $F(A) \to F(B)$ is an isomorphism. 
In addition, we require that if $A \cap S = \emptyset$ then $F(A) = \emptyset$.
The category of constructible cosheaves with natural transformations is denoted $\Cshc$.

The work of \cite{SilvaMunchPatel2014} gives the equivalence of categories $\Reeb  \equiv \Cshc$. 
In Figure~\ref{fig:geometric-roadmap}, when $d=1$, the functor $\CC:\Rdtop \to \Set^{\Open(\R^d)}$ (given in Figure~\ref{fig:roadmap}) restricts to a functor $\CC: \Rtopc \to \Cshc$.
Its further restriction $\CC: \Reeb \to \Cshc$ is exactly the functor used in \cite{SilvaMunchPatel2014} to give the equivalence of categories.  
In addition, $\CC$ has an ``inverse'' functor $\DD:\Cshc \to \Reeb$ which can turn a constructible cosheaf back into a geometric object through the display locale construction \cite{Woolf2013}.
This construction also satisfies the equality $\RR = \DD\CC$ due to the commutativity of the upper right triangle in Figure~\ref{fig:geometric-roadmap} (as proved in Section 3.5 of~\cite{SilvaMunchPatel2014}). Therefore constructing the (geometric) Reeb graph from well behaved data is the same as creating its categorical representation, and then turning it back into a geometric object. 

\subparagraph*{Our result.}
The above result implies that because we can turn \emph{any} constructible cosheaf back into a geometric Reeb graph, we can now turn the mapper, defined previously as a categorical object, back into a geometric object.  
In this spirit, let $\MM_K(\X,f) := \DD\PP_K\CC_K(\X,f)$ be the geometric representation of the mapper object, referred to as the \emph{geometric mapper} (following the rectangular diagram in Figure~\ref{fig:geometric-roadmap}), and let $\RR(\X,f)$ be the geometric Reeb graph.  
Then, the equivalence of categories gives us the following immediate corollary to Theorem~\ref{theorem:mapper-convergence}. 

\begin{corollary}
\label{theorem:geometric-convergence}
Given a constructible $\R$-space $(\X, f)$ with $f: \X \to \R$, let $\UU = \{U_\alpha\}_{\alpha \in A}$ be a good cover of $f(\X) \subseteq \R$, and let $K$ be the nerve of the cover.
Then 
$$ d_I(\RR(\X, f), \MM_K(\X, f)) \leq \mathrm{res}(\UU).$$
\end{corollary}
In particular, because the interleaving distance is an extended metric when $d=1$, this implies that a sequence of mappers for more refined covers $\UU$ converges to the Reeb graph geometrically. 
Recent work has also investigated this convergence problem using the bottleneck distance for the extended persistence diagrams~\cite{CarriereOudot2015}; instead, we use the interleaving distance.

\subparagraph*{Algorithm for geometric mapper.}
Constructing the geometric representation of $1$-dimensional mapper from its categorical representation follows a simple algorithm (as illustrated in Figure~\ref{fig:ReebGraph}).  
For the purpose of exposition, we assume that the mapper is constructed with a finite, connected, minimal cover (a cover with no subcover) and that the number of connected components over each cover element is finite.
We further assume that the open sets (intervals) in $\UU = \{U_i = (a_i,b_i)\}_{i=1}^n$ can be ordered and satisfy
$a_1 < a_2 < b_1 < a_3 < b_2 < \cdots <  a_{n-1} < b_{n-1} < b_n$.
For ease of notation, we assume there are extra intervals $U_0 = (a_0, b_0)$ with $a_0 <a_1 <b_0<b_1$ and $U_{n+1} = (a_{n+1}, b_{n+1})$ with $b_{n-1}<a_{n+1}<b_n<b_{n+1}$ and such that $f\inv(U_0) = f\inv(U_{n+1}) = \emptyset$.
Let $M:=M(\UU, f)$ be the mapper with the added property that for any cover element $U_i$, we store the vertices corresponding to connected components of $f\inv(U_i)$ in the  set $F(i)$.
Furthermore, let $M[i]$ be the subgraph of $M$ induced by the collection of vertices $F(i)$, and let $M[i,i+1]$ be the subgraph of $M$ induced by the vertices $F(i) \cup F(i+1)$.
Note that for any small enough interval $I \subset (a_{i+1},b_{i})$, the colimit construction for $I$ gives exactly the connected components over the union $U_i \cup U_{i+1}$, which is equivalently represented by the connected components of $M[i,i+1]$.
For any small enough interval $I \subset (b_{i-1},a_{i+1})$, the colimit construction for $I$ gives the connected components over $U_i$, and thus is represented by the connected components of $M[i]$, which are just the vertices.

Thus, the geometric mapper, $\MM_K(\Xspace,f) = (\Xspace', f')$, a graph $\Xspace'$ equipped with a function $f'$, can be constructed based on a combinatorial structure described below.
For each interval $[b_{i-1},a_{i+1}]$, add an edge $uv$ with two new pink vertices for each vertex in $M[i]$ (see Figure \ref{fig:ReebGraph} Algorithm). 
Set $f'(u) = b_{i-1}$ and set $f'(v) = a_{i+1}$.
For each interval $[ a_{i+1}, b_i]$, add an edge $wx$ with two new yellow vertices for each connected component in $M[i, i+1]$.
Set $f'(w) = a_{i+1}$ and $f'(x) = b_i$.
Now, we have a combinatorial structure which consists of a collection of disjoint edges spread across each of the intervals defined by the cover, and each edge has a top vertex and a bottom vertex given by the function values.
A pink and a yellow vertex are called equivalent if the vertex sets corresponding to them in $M[i]$ and $M[i,i+1]$ respectively have a nontrivial intersection.
The graph $\Xspace'$ resulting from identifying (i.e.~gluing) equivalent vertices with the same function value of $f'$  is the geometric mapper.
Such an algorithm relies on subroutines of union-find, therefore it inherits the complexity of union-find that varies depending on naive or advanced implementations.  

\section{Discussion}
\label{sec:discussion}

The authors of~\cite{CarriereOudot2015} asked whether it is possible to describe the mapper as a particular constructible cosheaf. We addressed this question for $d=1$ in Section \ref{sec:GeometricReps}: 
we described the mapper as a constructible cosheaf when it is passed to the continuous version.     
We suspect that our geometric results hold in the case $d > 1$.
That is, with the proper notion of constructibility for $\R^d$-spaces and cosheaves, we will have both an equivalence of categories, and a proof that the interleaving distance is an extended metric, not just a pseudometric; 
and therefore the mapper converges to the Reeb space on the space level. 
Our results are first steps towards providing a theoretical justification for the use of discrete objects (mapper and JCN) as approximations to the Reeb space with guarantees. 
Some future directions include creating categorical interpretation of multiscale mapper~\cite{DeyMemoliWang2015} and studying distance metrics between Jacobi sets in the categorical setting. 


\subparagraph*{Acknowledgements.}

BW would like to thank the support from NSF IIS-1513616.  



\bibliography{tda_vis-SoCG}

\begin{thebibliography}{10}

\bibitem{BubenikdeSilvaScott2014}
Peter Bubenik, Vin de~Silva, and Jonathan Scott.
\newblock Metrics for generalized persistence modules.
\newblock {\em Foundations of Computational Mathematics}, 15(6):1501--1531,
  2015.

\bibitem{CarrDuke2014}
Hamish Carr and David Duke.
\newblock Joint contour nets.
\newblock {\em {IEEE} Transactions on Visualization and Computer Graphics},
  20(8):1100--1113, 2014.

\bibitem{CarrSnoeyinkPanne2010}
Hamish Carr, Jack Snoeyink, and Michiel van~de Panne.
\newblock Flexible isosurfaces: Simplifying and displaying scalar topology
  using the contour tree.
\newblock {\em Computational Geometry}, 43:42--58, 2010.

\bibitem{CarriereOudot2015}
Mathieu Carri\'ere and Steve Oudot.
\newblock Structure and stability of the 1-dimensional mapper.
\newblock {\em Symposium on Computational Geometry (to appear);
  arXiv:1511.05823}, 2016.

\bibitem{ChazalSun2014}
Fr\'{e}d\'{e}ric Chazal and Jian Sun.
\newblock {G}romov-{H}ausdorff approximation of filament structure using
  {R}eeb-type graph.
\newblock {\em Proceedings 13th Annual Symposium on Computational Geometry},
  pages 491--500, 2014.

\bibitem{Curry2014}
Justin Curry.
\newblock {\em Sheaves, Cosheaves and Applications}.
\newblock PhD thesis, University of Pennsylvania, 2014.

\bibitem{SilvaMunchPatel2014}
Vin de~Silva, Elizabeth Munch, and Amit Patel.
\newblock Categorification of {R}eeb graphs.
\newblock {\em Discrete and Computational Geometry (to appear);
  arXiv:1501.04147}, 2016.

\bibitem{DeyMemoliWang2015}
Tamal~K. Dey, Facundo M\'{e}moli, and Yusu Wang.
\newblock Mutiscale mapper: A framework for topological summarization of data
  and maps.
\newblock arXiv:1504.03763, 2015.

\bibitem{EdelsbrunnerHarer2002}
Herbert Edelsbrunner and John Harer.
\newblock Jacobi sets of multiple {M}orse functions.
\newblock In F.~Cucker, R.~DeVore, P.~Olver, and E.~S\"{u}li, editors, {\em
  Foundations of Computational Mathematics, Minneapolis 2002}, pages 37--57.
  Cambridge University Press, 2002.

\bibitem{EdelsbrunnerHarerPatel2008}
Herbert Edelsbrunner, John Harer, and Amit~K. Patel.
\newblock Reeb spaces of piecewise linear mappings.
\newblock {\em Proceedings 24th Annual Symposium on Computational Geometry},
  pages 242--250, 2008.

\bibitem{HarveyWangWenger2010}
William Harvey, Yusu Wang, and Rephael Wenger.
\newblock A randomized ${O}(m\log{m})$ algorithm for computing {R}eeb graphs of
  arbitrary simplicial complexes.
\newblock {\em {ACM} Symposium on Computational Geometry}, pages 267--276,
  2010.

\bibitem{Hatcher2002}
Allen Hatcher.
\newblock {\em Algebraic Topology}.
\newblock Cambridge University Press, 2002.

\bibitem{Kozlov2008}
Dimitry Kozlov.
\newblock {\em Combinatorial Algebraic Topology}.
\newblock Springer, 2008.

\bibitem{LumSinghLehman2013}
P.~Y. Lum, G.~Singh, A.~Lehman, T.~Ishkanov, M.~Vejdemo-Johansson,
  M.~Alagappan, J.~Carlsson, and G.~Carlsson.
\newblock Extracting insights from the shape of complex data using topology.
\newblock {\em Scientific Reports}, 3, 2013.

\bibitem{MacLane1978}
Saunders Mac~Lane.
\newblock {\em Categories for the Working Mathematician}.
\newblock Springer-Verlag, New York, NY, 2nd edition, 1978.

\bibitem{MunchWang2016}
Elizabeth Munch and Bei Wang.
\newblock Convergence between categorical representations of {Reeb} space and
  mapper.
\newblock arXiv:1307.7760, 2016.

\bibitem{NicolauaLevinebCarlsson2011}
Monica Nicolaua, Arnold~J. Levineb, and Gunnar Carlsson.
\newblock Topology based data analysis identifies a subgroup of breast cancers
  with a unique mutational profile and excellent survival.
\newblock {\em Proceedings National Academy of Sciences of the United States of
  America}, 108(17):7265--7270, 2011.

\bibitem{Parsa2012}
Salman Parsa.
\newblock A deterministic ${O}(m\log{m})$ time algorithm for the {R}eeb graph.
\newblock {\em Proceedings 29th Annual Symposium on Computational Geometry},
  pages 269--276, 2012.

\bibitem{Patel2010}
Amit Patel.
\newblock {\em Reeb Spaces and the Robustness of Preimages}.
\newblock PhD thesis, Duke University, 2010.

\bibitem{Penna1978}
Michael~A. Penna.
\newblock On the geometry of combinatorial manifolds.
\newblock {\em Pacific Journal of Mathematics}, 77(2):499--522, 1978.

\bibitem{Reeb1946}
G.~Reeb.
\newblock Sur les points singuliers d'une forme de pfaff completement
  intergrable ou d'une fonction numerique [on the singular points of a complete
  integral pfaff form or of a numerical function].
\newblock {\em Comptes Rendus Acad.Science Paris}, 222:847--849, 1946.

\bibitem{SinghMemoliCarlsson2007}
Gurjeet Singh, Facundo M\'emoli, and Gunnar Carlsson.
\newblock Topological methods for the analysis of high dimensional data sets
  and {3D} object recognition.
\newblock In {\em Eurographics Symposium on Point-Based Graphics}, 2007.

\bibitem{Stovner2012}
Roar~Bakken Stovner.
\newblock On the mapper algorithm: A study of a new topological method for data
  analysis.
\newblock Master's thesis, Norwegian University of Science and Technology,
  2012.

\bibitem{Woolf2013}
Jon Woolf.
\newblock The fundamental category of a stratified space.
\newblock arXiv:0811.2580, 2013.

\end{thebibliography}


\end{document}